\documentclass[aps,prd,twocolumn,groupedaddress,eqsecnum]{revtex4}
\input epsf

\begin{document}


\title{QCD String Structure in Vector Confinement} 
\author{Theodore J. Allen}
\affiliation{Physics Department, Hobart \& William Smith Colleges \\
Geneva, New York 14456 USA}

\author{Todd Coleman}
\author{M. G. Olsson}
\affiliation{Department of Physics, University of Wisconsin, \\
1150 University Avenue, Madison, Wisconsin 53706 USA }

\author{Sini\v{s}a Veseli}
\affiliation{Fermi National Accelerator Laboratory \\
P.O. Box 500, Batavia, Illinois 60510 USA}

\date{\today}

\begin{abstract}
We demonstrate that the Nambu-Goto string spectroscopy with massless quarks
is replicated in highly excited states of the linear vector confinement
potential.  For deep radial excitations we observe that the Regge slope,
spacing between daughter trajectories, and absolute state energies agree
with those of the QCD string.
\end{abstract}
\maketitle

\section{Introduction}\label{sec:intro}

We present here an interesting simple observation on electric potential
confinement.  Our conclusions concern the nature of potential models and
their relationship to QCD string confinement.  In the case of the time
component vector (TCV) potential we note a remarkable transition in the
Regge spectroscopy from the leading trajectories with slopes $1/4a$ to deep
daughter slopes of $1/\pi a$.  In fact, the exact string spectroscopy is
found to be embedded in this regime.  The latter slope is characteristic of
the Nambu-Goto QCD string with one fixed end \cite{ref:one}. We first
discuss some general properties of the spinless Salpeter (SS) or square
root equation.  For the linear TCV potential we then compute the Regge
structure in the orbitally and radially dominant regimes.

\section{Spinless Salpeter Equation}\label{sec:salpeter}

The spinless Salpeter equation has long been attractive as a minimal
generalization of the non-relativistic Schr\"odinger equations with
relativistic kinematics.  Its main difficulty is that it involves the
square root of a differential operator, sometimes called a
pseudo-differential operator, which is difficult to treat analytically.
Even so, it is easily implemented numerically.  The introduction of a
variational method \cite{ref:two} allowed straightforward numerical
solutions involving square roots and even transcendental functions of
operators.  In this section we discuss, though do not prove, the necessity
of using the spinless Salpeter equation in place of a Klein-Gordon type
equation when dealing with a pure vector potential.

\begin{figure}[hbp]
\epsfxsize = \linewidth 
\hspace*{-5mm}
\epsfbox{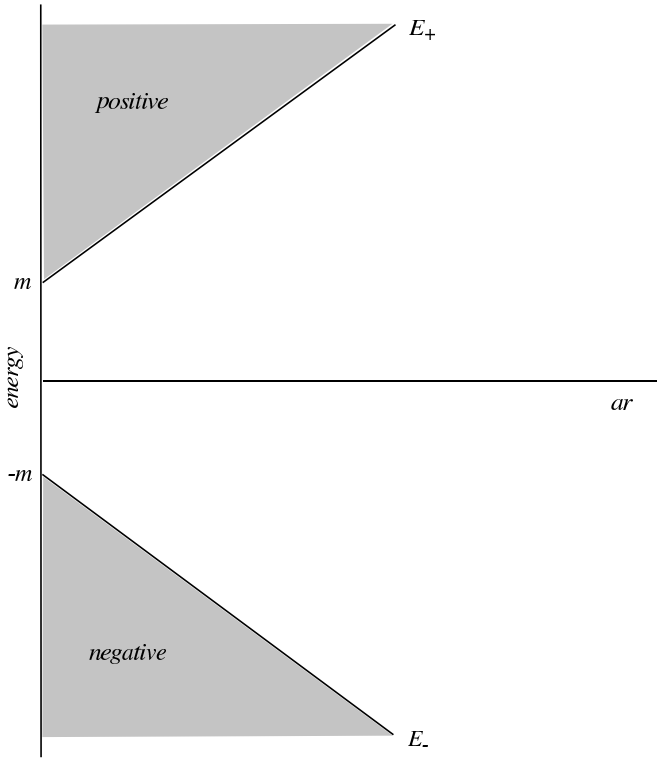}
\vskip 1 cm
\caption{S-wave classical turning points for Eq.~(\protect\ref{Eturning}) with scalar confinement.}
\label{fig:one}
\end{figure}

\begin{figure}[hbp]
\epsfxsize = \linewidth 
\hspace*{-5mm}
\epsfbox{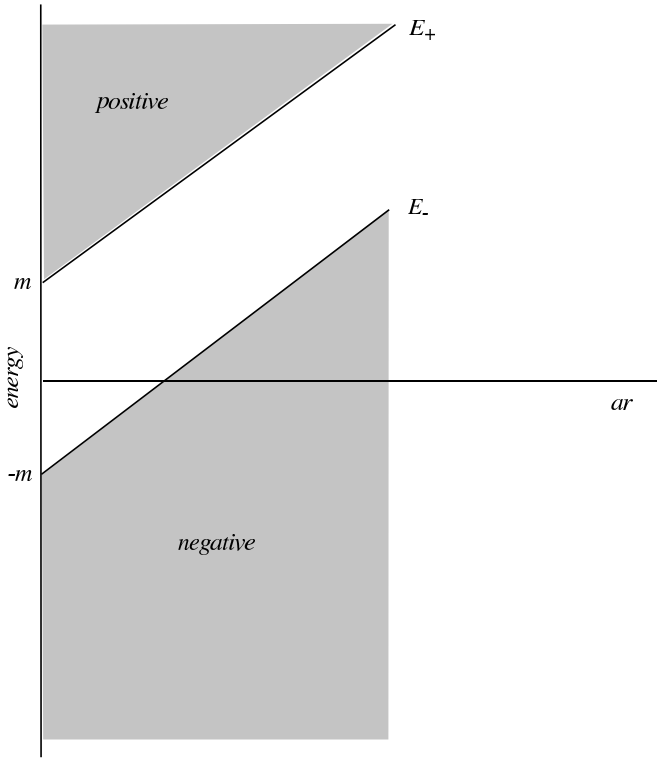}
\vskip 1 cm
\caption{S-wave classical turning points for Eq.~(\protect\ref{Eturning}) with vector confinement.}
\label{fig:two}
\end{figure}

Starting from the simple classical Lagrangian,
\begin{equation}\label{lagrangian}
L = -(m + S(r))\sqrt{1-v^2} - V(r) , 
\end{equation}
describing a particle of mass $m$ moving with velocity $v$ in a Lorentz
scalar potential $S(r)$ and a Lorentz time-component vector potential
$V(r)$, we find that the Hamiltonian,
\begin{equation}\label{hamiltonian}
H = \pm\sqrt{p^2 + (m+S(r))^2} + V(r) ,
\end{equation}
has a sign ambiguity.  This ambiguity arises from the elimination of
the velocity in favor of the momentum
\begin{equation}
 {\bf p} = {\partial L\over \partial {\bf v}} = (m + S(r))\, \gamma\,
  {\bf v} , 
\end{equation}
where $\gamma = (1-v^2)^{-1/2}$. 
The positive sign in Eq.~(\ref{hamiltonian}) corresponds to the usual
positive energy states and yields the correct non-relativistic limit
\begin{equation}
H \buildrel {|p| \ll m} \over \longrightarrow m + {p^2\over2m} + S(r) + V(r) .
\end{equation}
The squared form of Eq.~(\ref{hamiltonian}),
\begin{equation}\label{GKG}
(H - V(r))^2 = p^2 + (m+S(r))^2 ,
\end{equation}
known as the generalized Klein-Gordon (GKG) equation, behaves quite
differently in the cases of a pure vector and a pure scalar potential.
This difference is most easily seen from consideration of the s-wave
turning points, where $p^2 = 0$.  Eq.~(\ref{hamiltonian}) determines the
energy as a function of the potentials at the turning points,
\begin{equation}\label{Eturning}
E = \pm (m + S) + V .
\end{equation}
We show this turning point condition separately for pure scalar and pure
vector linear confining potentials in Figs.~\ref{fig:one} and
\ref{fig:two}.  In Fig.\ \ref{fig:one} we see that for the pure scalar case
there is no overlap of the positive and negative branches of
Eq.~(\ref{Eturning}).  In Fig.\ \ref{fig:two}, on the other hand, we see
that for any $E>m$ in pure TCV linear confinement there are states on both
positive and negative branches.  A completely analogous observation
\cite{ref:three} has been made earlier in the context of the Dirac
equation.

Since the GKG equation (\ref{GKG}) is quadratic and contains both branches of
Eq.~(\ref{hamiltonian}), it can have no normalizable solutions in pure TCV
confinement because there is tunneling to the negative energy states, which
becomes catastrophic for small quark mass. This difficulty is the
well-known ``Klein Paradox'' \cite{ref:four}. On the other hand, the GKG
equation with pure scalar confinement does have well defined normalizable
solutions because there is no possibility of tunneling to the negative
branch.

The spirit of the Salpeter equation \cite{ref:five} is to conserve a
definite particle number.  To this end, energy projection operators are
employed to remove the negative energy spectrum and the fermion TCV
Salpeter equation has normalizable solutions for confined
quarks \cite{ref:three}.  The spinless Salpeter equation for TCV confinement
achieves the same end more simply by explicitly including only the positive
branch of Eq.~(\ref{hamiltonian}).  For p-wave and higher angular momentum
states, we consider the Hamiltonian in spherical coordinates,
\begin{equation}\label{eq:six}
H = \sqrt{p_r^2 + {J^2\over r^2} + m^2} + ar ,
\end{equation}
which we obtain from dividing the momentum into radial and angular parts.
The wave equation to be solved is then the eigenvalue problem 
\begin{equation}\label{eq:five}
H|\psi\rangle = E|\psi\rangle 
\end{equation}
for the pseudo-differential operator given by Eq.~(\ref{eq:six}) with the
usual operator replacements.


\section{Semi-Classical Analytic Quantization}

We have discussed the necessity of using the SS equation instead of the GKG
equation with vector confinement. In semi-classical language, the necessity
of using the SS equation becomes the necessity of using the turning points
given by Eq.~(\ref{Eturning}) with the positive sign only.  In this
section we calculate the semi-classical spectrum of the SS equation with
TCV confinement.  We first rewrite the SS equation (\ref{eq:six}) in terms
of dimensionless variables
\begin{eqnarray}\label{dimless}
x & \equiv & {ar\over E} ,\nonumber \\
\beta & \equiv & {aJ \over E^2}, \\
P_r & \equiv & {p_r\over E}. \nonumber 
\end{eqnarray}
In the case of a massless quark, we have
\begin{eqnarray}\label{poly}
P_r & = & {\sqrt{Q}\over x} , \\
Q & \equiv & (x- x_{1-})(x - x_{2-}) ( x_{1+} - x) (x_{2+} - x) , \nonumber 
\end{eqnarray}
where
\begin{eqnarray}\label{roots}
x_{1\pm} & = & \frac12\left(1\pm\sqrt{1 - 4\beta}\right) , \nonumber \\
x_{2\pm} & = & \frac12\left(1\pm\sqrt{1 + 4\beta}\right) .
\end{eqnarray}
The dimensionless physical turning points resulting from the SS equation
are $x_{1\pm}$. The dimensionless constants $x_{2\pm}$ are turning points
of the GKG equation, but not of the SS equation.

In the regime we are investigating the angular momentum $J$ is limited but
the energy $E$ is large. It follows that $\beta \ll 1$.  To see this, we
note that even for the leading (circular orbit) Regge trajectory $J/E^2 =
1/4a$ (see the appendix) and hence $\beta_{\rm leading} = 1/4$.  For fixed
$J$ and large $E$, $\beta$ becomes small.  In this limit the leading
behavior of the roots (\ref{roots}) is
\begin{equation}
\begin{tabular}{lll}
& $x_{1-}  =  \beta + \beta^2 + \ldots ,$  & $x_{1+} = 1 - \beta - \beta^2 -
\ldots ,$ \\
& $x_{2-}  = - \beta + \beta^2 + \ldots ,$ & $x_{2+} = 1 + \beta - \beta^2
+ \ldots .$ \\
\end{tabular}
\end{equation}

The SS equation may be quantized semi-classically by the usual method
\cite{ref:six}
\begin{equation}\label{eq:semi1}
\int_{r_-}^{r_+} dr\, p_r  = \pi\left(n+\frac12\right) .
\end{equation}
In dimensionless variables (\ref{dimless}), the quantization condition
(\ref{eq:semi1}) becomes
\begin{equation}\label{semi}
\int_{x_-}^{x_+} dx\, P_r(x) = {a\pi\over E^2}\left(n+\frac12\right) .
\end{equation}
The TCV quantization is achieved by integrating Eq.~(\ref{poly}).  The
method is quite accurate for all states and becomes exact for states with
many nodes in the radial wavefunction.  This is the regime we are
particularly interested in.

Although the integral of Eqs.~(\ref{poly}), (\ref{semi}) can be expressed in
terms of elliptic integrals, it is more efficient to approximate the
integrand first for small $\beta$.  The approximation
\begin{equation}\label{approxintegrand}
{\sqrt{Q}\over x} \simeq {\sqrt{(x-x_{1-})(x-x_{2-})}\over x}\left[1-x -
{\beta^2\over 2(1-x)} - \beta^2\right] 
\end{equation}
reproduces the exact integrand to better than $0.5\%$ throughout most of
the region of integration, even when $\beta=0.1$.  The accuracy of this
approximation is shown graphically in Fig.~3, where we plot both the exact
integrand (\ref{poly}) and the approximation (\ref{approxintegrand}) for
$\beta=0.1$.

With the approximation (\ref{approxintegrand}), the quantization integral
(\ref{semi}) is relatively easily evaluated.  To order $\beta$ we find
\begin{equation}
\int_{x_-}^{x_+} {\sqrt{Q}\over x} \, dx \simeq \frac12 - {\pi\beta\over 2}
+ {\cal O}(\beta^2),
\end{equation}
which  immediately leads to the result 
\begin{equation}
{E^2\over \pi a} = J + 2n + 1 .
\end{equation}
Upon making the Langer \cite{ref:seven} correction to take into account the
centrifugal singularity,
\begin{equation}
J\rightarrow J+\frac12 ,
\end{equation}
where $J$ is now the angular momentum quantum number, we find the final
spectroscopic relation
\begin{equation}\label{stringregge}
{E^2\over \pi a} = J + 2n + \frac32 .
\end{equation}

The relation (\ref{stringregge}) is identical to an analytic approximation
for highly radially excited QCD string, as well as being an excellent
approximation to its exact numerical solution \cite{ref:eight}.


\section{Comparison to Exact Numerical Results}

We can nail down our central result that the QCD string spectrum
(\ref{stringregge}) is replicated in the radially dominant regime of the
TCV potential by solving the TCV SS equation exactly numerically.  The
variational (Galerkin) method is well suited for solving eigenvalue
equations with mixed coordinate and momentum operators.  Briefly, the
method begins with a complete set of orthogonal states that can be Fourier
transformed.  The wavefunction is approximated as a superposition of the
lowest $N$ of those states.  The wave equation transforms into an $N\times
N$ matrix equation which is then diagonalized.  The accuracy of the
resulting eigenvalues and wavefunctions is measured by the dependence on
$N$ and the dependence of the scale parameter of the basis set.  Details
can be found in \cite{ref:nine}.  Some further considerations of the accuracy
of the variational method are presented in \cite{ref:ten}.

The TCV SS equation (\ref{eq:six}) with $m=0$ can thus be solved for a
variety of angular momenta, $J=0,1,2,\ldots$ and radial states,
$n=0,1,2,\ldots$.  The result is depicted in Fig.~4.  Plotted on this
figure are both the QCD string slope $1/\pi a$ (solid) and the usual
leading trajectory TCV slope $1/4a$ (dashed).  In both cases we start the
lines at $J=0$.  Although the leading trajectory ($n=0$) agrees with the
dashed prediction, the situation changes as we examine the higher radial
excitations.  At the larger $n$ values (deep daughters) we observe that the
solid line corresponding to the QCD string slope becomes accurate.

In Table~1 we provide the exact and the string solutions for the s-wave
radial excitations.  We see now that all aspects of the QCD string result
(\ref{stringregge}) work quite well.  The difference between excitations is
accurately two units of $E^2/\pi a$ and the absolute values of the state
energies correspond well to the QCD string to three significant figures.

\section{Summary and Discussion}

On occasion, some simple results are quite unexpected.  We did not expect to
find the $m=0$ QCD string spectroscopy in a potential model.  To summarize
our finding, we have examined both numerically and analytically the state
spectroscopy for a massless quark moving in the time-component vector
linear confinement potential.  For states with angular momentum much less
than the number of nodes we find the spectrum is exactly that of the QCD
string.  The Regge slope, radial excitation energy, and the absolute values
of the energy (\ref{stringregge}) are exactly what one expects from the QCD
string.

Simple results, even surprising ones, usually have simple explanations.  As
we have recently pointed out \cite{ref:eight}, the QCD string equations
reduce to the spinless Salpeter equation with a linear time-component
vector potential for the s-wave states.  Thus there is a natural physical
connection between the two systems.  We can understand this connection
simply.  A QCD string corresponds to a constant chromoelectric field in the
quark rest frame.  In the limit that the quark is moving radially, the QCD
string has no angular momentum.  The total energy of the system becomes
\begin{equation}
E_{\rm string} = \sqrt{{\bf p}^2 + m^2} + ar {\arcsin v_\perp\over v_\perp}
\buildrel {v_\perp \rightarrow 0} \over \longrightarrow  \sqrt{{p_r}^2 + m^2} + ar .
\end{equation}
In a potential model the field never carries angular momentum.  The
potential energy of the quark is linear in the distance from the origin,
which is the same as the energy of the string as long as the string is moving
radially so that there are no relativistic corrections due to its
transverse motion.  The total energy of the quark in a TCV potential,
\begin{equation}
E_{\rm TCV} = \sqrt{{\bf p}^2 + m^2} + ar,
\end{equation}
in the limit of vanishing $v_\perp$ becomes
\begin{equation}
\lim_{v_\perp \rightarrow 0} E_{\rm string} = \lim_{v_\perp \rightarrow 0}
E_{\rm TCV} = \sqrt{{p_r}^2 + m^2} + ar .
\end{equation}
It is thus natural to expect agreement between string
and TCV confinement in the radial dominant regime.

The real puzzle is why the Regge slope of the QCD string should be the
same for both circular and radial motions.  On one hand it is well known
from Nambu-Goto days that the Regge slope of a rotating QCD string with one
end fixed is $1/\pi a$.  We have seen that the TCV potential Regge slope
varies from $1/4a$ to $1/\pi a$ as one goes from orbital to radial motion
[see Fig.~4]. There is thus no obvious reason why this should not happen
with the QCD string.  Since the QCD string/TCV potential Regge slopes
coincide at $1/\pi a$ for nearly radial motions this explains the
remarkable uniform Regge structure seen previously from numerical solutions
of the QCD string \cite{ref:eight}.

\section*{Acknowledgments}
This work was supported in part by the US Department of Energy under
Contract No.~DE-FG02-95ER40896.

\appendix
\section{Nearly Circular Orbits}

For completeness we semi-classically quantize the TCV equation for nearly
circular orbits ($J \gg n$). Using the notation of
Eqs.~(\ref{dimless}--\ref{roots}), we consider the orbitally dominant regime.
In this case the turning points $x_{1+}$ and $x_{1-}$ are nearly equal.  By
Eq.~(\ref{roots}), this occurs exactly at
\begin{eqnarray}
x_c & = & \frac12 , \nonumber \\
\beta_c & = & \frac14 .
\end{eqnarray}
For nearly circular orbits we can expand in $\beta$ and $x$, giving
\begin{eqnarray}
x_{1\pm} & = & \frac12 \pm \sqrt{\frac14 - \beta} \nonumber \\
Q & \simeq & \frac1{16} - \beta^2 - \frac12\left(x-\frac12\right)^2 \\
  & \simeq & \frac12(x - x_{1-})(x_{1+} - x) \nonumber
\end{eqnarray}

Using the quantization condition (\ref{semi}), we obtain
\begin{equation}
{1\over 2\sqrt2}\left(1-2\sqrt\beta\right) = {a\over
E^2}\left(n+\frac12\right) ,
\end{equation}
with $\beta = a J/E^2 $.  Solving for $\sqrt\beta$, squaring and dropping
the small squared $n/E^2$ term, we obtain
\begin{equation}
{E^2\over 4a} = J + \sqrt2 n + \frac1{\sqrt2} .
\end{equation}
Finally, we make the Langer \cite{ref:seven} correction $J \rightarrow J +
\frac12$ and find \cite{ref:eight}
\begin{equation}
{E^2\over 4a} = J + \sqrt2 n + \frac12 + \frac1{\sqrt2} .
\end{equation}

\setbox1=\hbox{  
\begin{tabular}{|r|r|r|}
\hline
$n$ & Exact $\left({E^2\over \pi a}\right)$& $ 2n + \frac32$ \\ \hline \hline
0 & 1.59 & 1.50 \\
1 & 3.53 & 3.50 \\
2 & 5.52 & 5.50 \\ 
3 & 7.51 & 7.50 \\ 
4 & 9.51 & 9.50 \\
5 & 11.51 & 11.50 \\
6 & 13.51 & 13.50 \\
7 & 15.51 & 15.50 \\
8 & 17.51 & 17.50 \\
9 & 19.51 & 19.50 \\
10 & 21.51 & 21.50 \\
\hline
\end{tabular}
}
\begin{table}[hb]
\caption{ An exact numerical solution for s-wave TCV confinement compared
with the analytic approximation to the TCV/string showing the rapid
convergence of our WKB approximation as the number of radial nodes, $n$,
increases. }
\hfil\box1\hfil
\end{table}

\newpage

\newpage
\vspace{2cm}

\newpage
\vspace{2cm}

\begin{figure}[hbp]
\epsfxsize = \linewidth 
\hspace*{-5mm}
\epsfbox{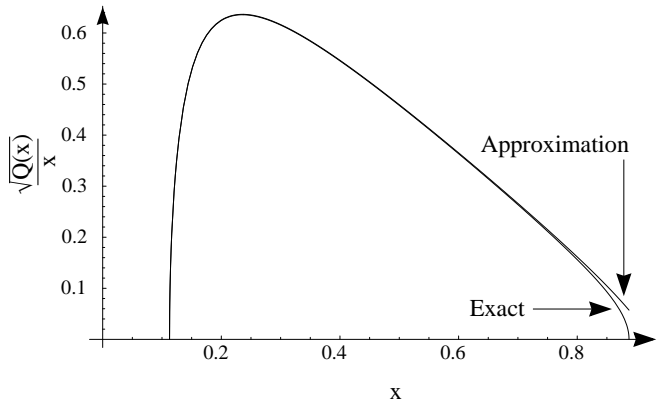}
\vskip 1 cm
\caption{The exact Bohr-Sommerfeld integrand (\protect\ref{poly}) and the
approximation (\protect\ref{approxintegrand}) for $\beta=0.1$. }
\label{fig:three}
\end{figure}
\newpage
\vspace{2cm}

\begin{figure}[hbp]
\epsfxsize = \linewidth 
\hspace*{-5mm}
\epsfbox{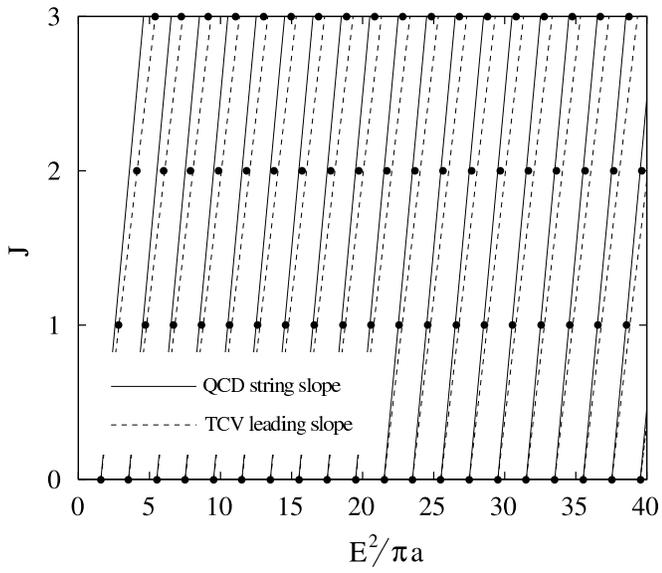}
\vskip 1 cm
\caption{Regge diagram of exact numerical solutions to
Eq.~(\protect\ref{eq:five}) with linear vector confinement (dots). The
solid lines are the QCD string trajectories with slope $1/\pi a$ and the
dashed lines have slope $1/4a$. We observe the transition from the leading
TCV slope to string slope.}
\label{fig:four}
\end{figure}
\newpage
\vspace{2cm}

\end{document}